# 2D SnS: a phosphorene analogue with strong in-plane electronic anisotropy

*Zhen Tian,[1,2,3] Chenglei Guo,[1,2,3] Mingxing Zhao,[2,3] Ranran Li,[2,3] and Jiamin Xue[1,2,3,4],* *

[1]Shanghai Institute of Optics and Fine Mechanics, Chinese Academy of Sciences, Shanghai 201800, China

[2]School of Physical Science and Technology, ShanghaiTech University, Shanghai 201210, China

[3]University of Chinese Academy of Sciences, Beijing 100049, China

[4]Center for Excellence in Superconducting Electronics (CENSE), Chinese Academy of Sciences, Shanghai 200050, China

ABSTRACT

We study the anisotropic electronic properties of 2D SnS, an analogue of phosphorene, grown by physical vapor transport. With transmission electron microscope and polarized Raman spectroscopy, we identify the zigzag and armchair directions of the as-grown 2D crystals. 2D SnS field-effect transistors with a cross-Hall-bar structure are fabricated. They show heavily hole-doped (~$10^{19}$ cm$^{-3}$) conductivity with strong in-plane anisotropy. At room temperature the mobility along the zigzag direction exceeds 20 cm$^2$V$^{-1}$s$^{-1}$, which can be up to 1.7 times of that in the armchair direction. This strong anisotropy is then explained by the effective-mass ratio along the two directions and agrees well with previous theoretical predictions. Temperature-dependent carrier density is used to find out the acceptor energy level to be ~45 meV above the valence band maximum. This value matches with a calculated defect level of 42 meV for Sn vacancies, indicating that Sn deficiency is the main cause of the p-type conductivity.

Phosphorene, a 2D format of black phosphorus, has become a star in the 2D material family.[1-3] Its most exotic property is arguably the in-plane anisotropic response to external stimulations, such as polarized light, electric field, strain, and so on.[3-7] This anisotropy has its origin rooted in the puckered lattice structure, and provides a new degree of freedom to explore in the 2D materials. Here we report that phosphorene is not alone by showing that a phosphorene analogue, 2D SnS, also poses interesting anisotropic properties. Bulk SnS has similar layered structure as black phosphorus, and has been extensively studied as an earth-abundant thin-film solar energy absorption material.[8-10] Recently, its 2D properties have attracted great theoretical interests. Many exotic phenomena have been predicted, including valley-dependent transport excited by linearly polarized light,[11] giant piezoelectricity,[12] reversible structure switched by strain or electric field,[13] multiferroicity,[14] and so on. However, experimental study is very scarce. In this article, we focus on the electronic transport properties of physical vapor transport (PVT) grown few-layer SnS, with thicknesses between 12 nm (~20 layers) and 14 nm (~25 layers). We find that the SnS nanoplates are heavily p doped (~$10^{19}$ cm$^{-3}$), and the on-off ratio can be up to $10^4$ at 77 K. With the help of transmission electron microscope (TEM) and polarized Raman spectroscopy, we identify the different crystal axes of the as-grown SnS nanoplates. Field-effect transistors (FETs) with a novel cross-Hall-bar structure are fabricated to study the intrinsic anisotropic transport properties. At room temperature, the mobility along the zigzag direction exceeds 20 cm$^2$V$^{-1}$s$^{-1}$, a respectable value for a 2D metal chalcogenide;[15] and is much higher than that along the armchair direction. The ratio of the two values reaches up to 1.7, comparable to that of phosphorene;[7] but interestingly the order is switched, meaning that in phosphorene the armchair direction has the higher mobility. This anisotropy comes from the different dispersion relations and hence effective masses along the two crystal directions. To identify the source of the heavy hole doping, carrier density as a function of temperature is carefully analyzed and the activation energy of acceptors is found to be ~45 meV, which shows quantitative agreement with calculated energy level of Sn vacancies.[16] Our study of 2D SnS paves the way for future exploration of its many predicted extraordinary properties.

RESULTS AND DISCUSSION

The 2D SnS nanoplates are synthesized with the PVT method, similar to the growth of SnSe.[17] A schematic of the PVT system is shown in Figure 1a. The SnS powder is heated in an evacuated two-zone tube furnace. SnS vapor is carried downstream with high purity Ar gas and deposited onto mica surfaces (see the Methods for more growth details). After the growth, a large number of 2D SnS nanoplates are formed on the surface of mica, as shown in the optical microscope (OM) images of Figure 1b. It can be observed that the as-synthesized SnS have approximate rhombic shape with lateral dimensions of 5 to 15 μm and clear facets, indicating the single crystallinity of individual nanoplates. Unlike $SiO_2$/Si supported 2D materials, whose thickness can be judged from the interference color, SnS nanoplates with different thicknesses on mica have low color contrast under reflection-mode microscope (Figure 1b upper part). Due to the high absorption coefficient[18, 19] (~$10^5$ cm$^{-1}$) of SnS in the visible light range, we found that the thickness of SnS on mica can be better identified with transmission-mode microscopy (Figure 1b lower part). Different transparencies can be observed corresponding to different thicknesses. For a more accurate measurement and device fabrication, we need to transfer the SnS nanoplates from mica to other substrates. One common method is to spin coat the sample with some polymer (e.g. PMMA) and then etch away the substrate to harvest the nanoplates.[20-22] However, chemical etching would unavoidably leave behind residues on the bottom of the nanoplates and degrade their electronic quality. Inspired by the water-mediated transfer method of nanostructures,[23, 24] we exploit the hydrophobic-hydrophilic properties of the interfaces, and use deionized water to separate PMMA with SnS from the mica substrate. This method gives clean SnS nanoplates with $SiO_2$/Si substrate as evidenced by the AFM (Asylum research, MFP-3D) images in Figures 1c and 1d. With the AFM we find that the minimum thickness of as grown SnS is about 6 nm (~10 layers). Thicker nanoplates generally have well-defined straight edges (Figure 1c) compared with the thinner ones (Figure 1d), which indicates a better crystallinity of the former. Growing even thinner SnS (down to monolayer) with good quality is a subject of future study.

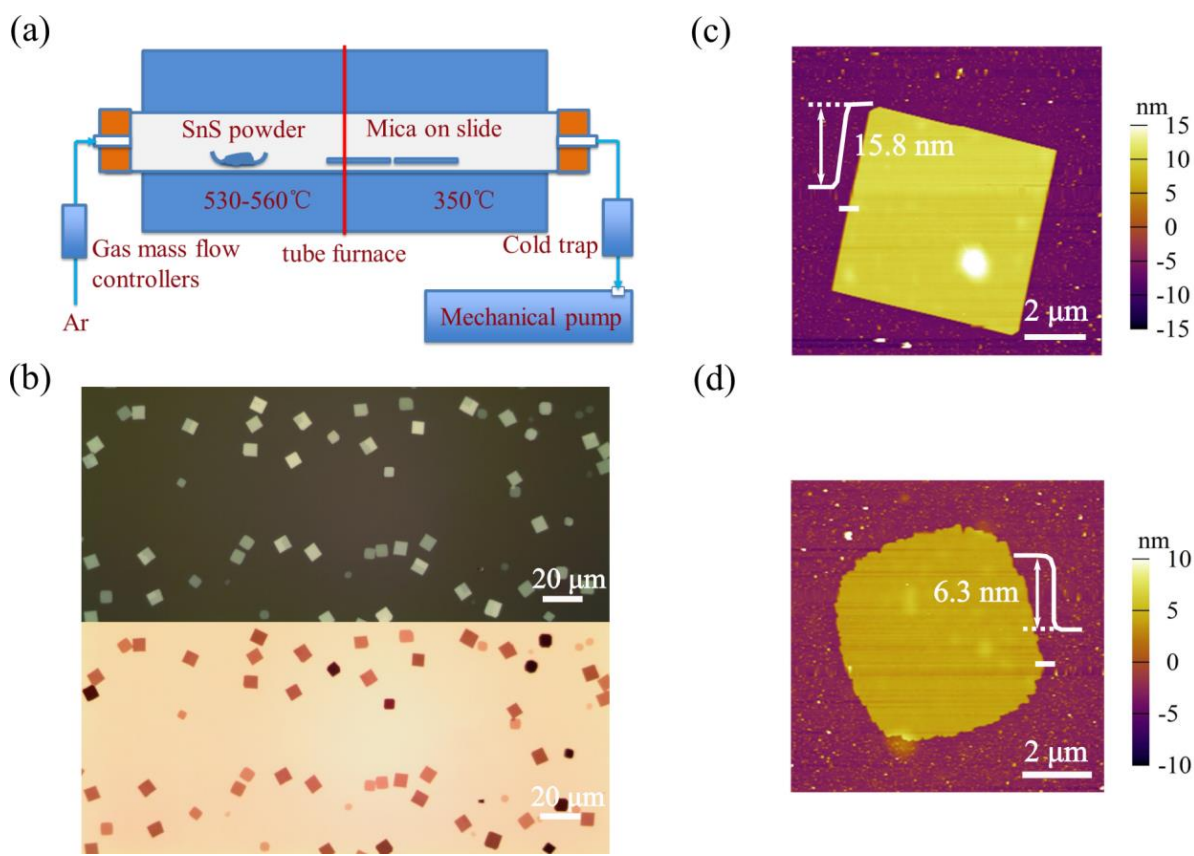

**Figure 1.** (a) Schematic of the PVT growth system. The 2D SnS is grown on mica. (b) OM images of SnS nanoplates on mica after growth. The upper image is obtained with reflection-

mode microscopy and the lower one transmission mode. (c, d) AFM images of 15.8 nm and 6.3 nm 2D SnS nanoplates transferred onto $SiO_2$ substrates.

Bulk SnS has a puckered structure similar to that of black phosphorus[25, 26] (see the atomic model in Figure 2a). It belongs to the orthorhombic crystal system with the space group Pnam (62) when a coordinate system as defined in Figure 2a is used. Conventionally, b axis is named as the armchair direction and c axis denotes the zigzag direction. To compare the structure of the synthesized 2D nanoplates with bulk SnS, transmission electron microscope (TEM, TECNAI, $G^2$S-TWIN) was used to study the nanoplates transferred onto TEM grids. A high-resolution TEM (HRTEM) image of a representative nanoplate is shown in Figure 2b with corresponding selected-area electron diffraction (SAED) pattern in Figure 2c. The perfect rhombus lattice fringes in Figure 2b and the sharp SAED spots in Figure 2c confirm the single-crystallinity of the nanoplates. The lattice spacing obtained from Figures 2b and 2c match well with that of the bulk SnS[27-29] (see the figure caption for details), showing no structural change from bulk to few-layer nanoplates. It is noted that in the electron diffraction pattern of Figure 2c, the spots corresponding to planes (001) and (010) (and their equivalents) are missing, due to the systematic absence.

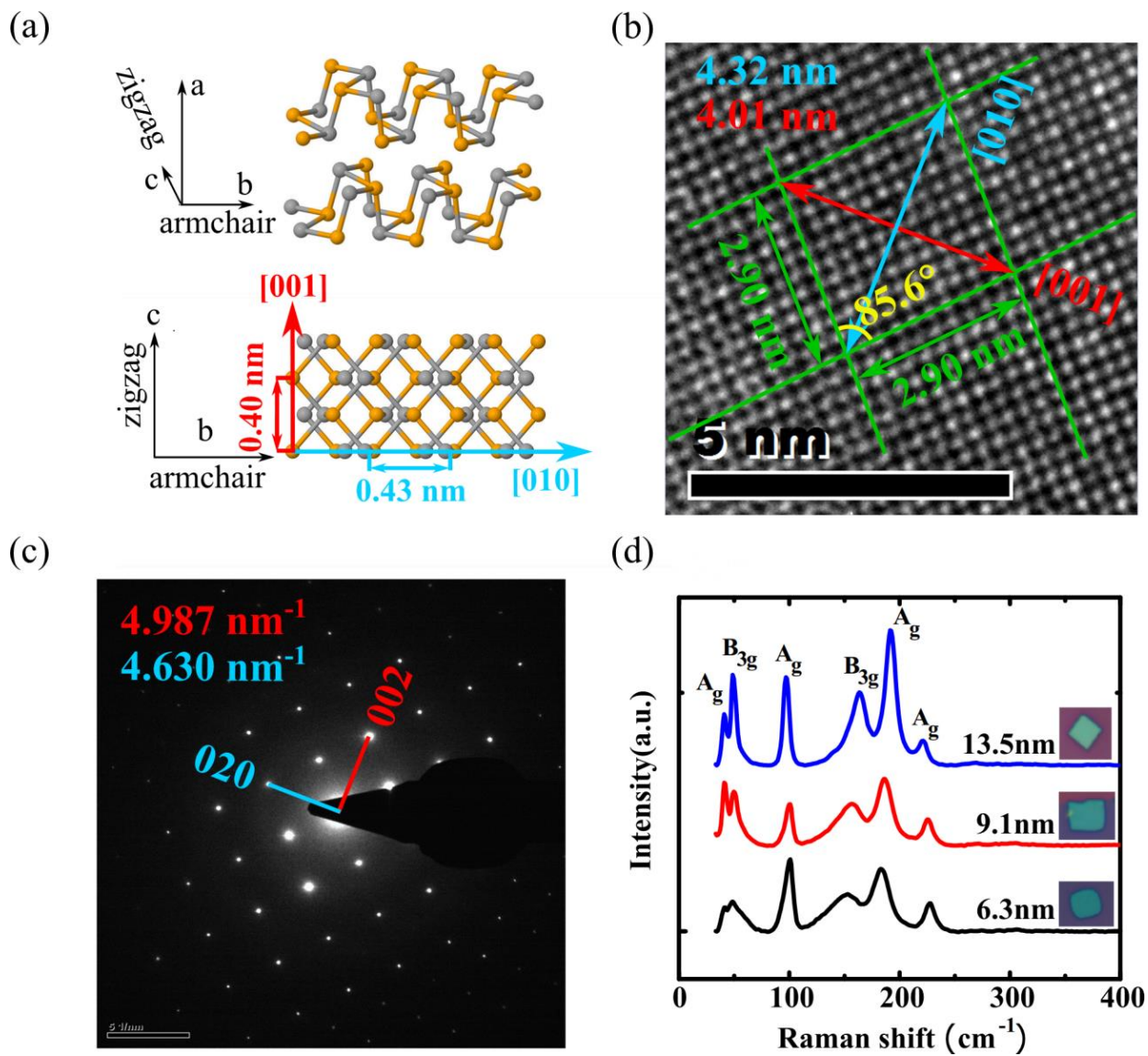

**Figure 2.** (a) Atomic model of bulk SnS crystal structure viewed with 3D perspective (upper) or along the interlayer direction (lower). (b) HRTEM image of an as-grown 2D SnS. The lengths of ten lattice fringes along the [011], [01$\bar{1}$], [010] and [001] directions are 2.90 nm, 2.90 nm, 4.32 nm and 4.01 nm, respectively. These values match well with those of the bulk as shown in (a). Color coding is used for clearance. (c) The SAED pattern corresponding to (b). Distances between the (020) and (002) crystal planes are obtained by measuring the lengths of the cyan and red lines, respectively. (d) Raman spectra of as-grown 2D SnS nanoplates with different thicknesses. Insets are the corresponding OM images.

Raman spectroscopy was also performed to provide more information about the quality of the as-grown SnS nanoplates with different thicknesses. Figure 2d shows the Raman spectra of three SnS nanoplates with the thicknesses of 6.3 nm, 9.1 nm and 13.5 nm. The peak positions ($A_g$ modes[30] at 40.0 cm$^{-1}$, 97.2 cm$^{-1}$, 191.9 cm$^{-1}$ and 218.7 cm$^{-1}$; and $B_{3g}$ modes at 49.4 cm$^{-1}$ and 163.2 cm$^{-1}$) match very well with previous study on bulk single crystalline SnS.[31] It can be seen from the data that the thicker nanoplate has stronger and sharper Raman peaks indicating a better crystal quality, consistent with the nanoplate-edge morphologies obtained by the AFM in Figure 1. Throughout the rest of this study, we focus on the nanoplates with well-developed shapes and thicknesses of 12 to 14 nm.

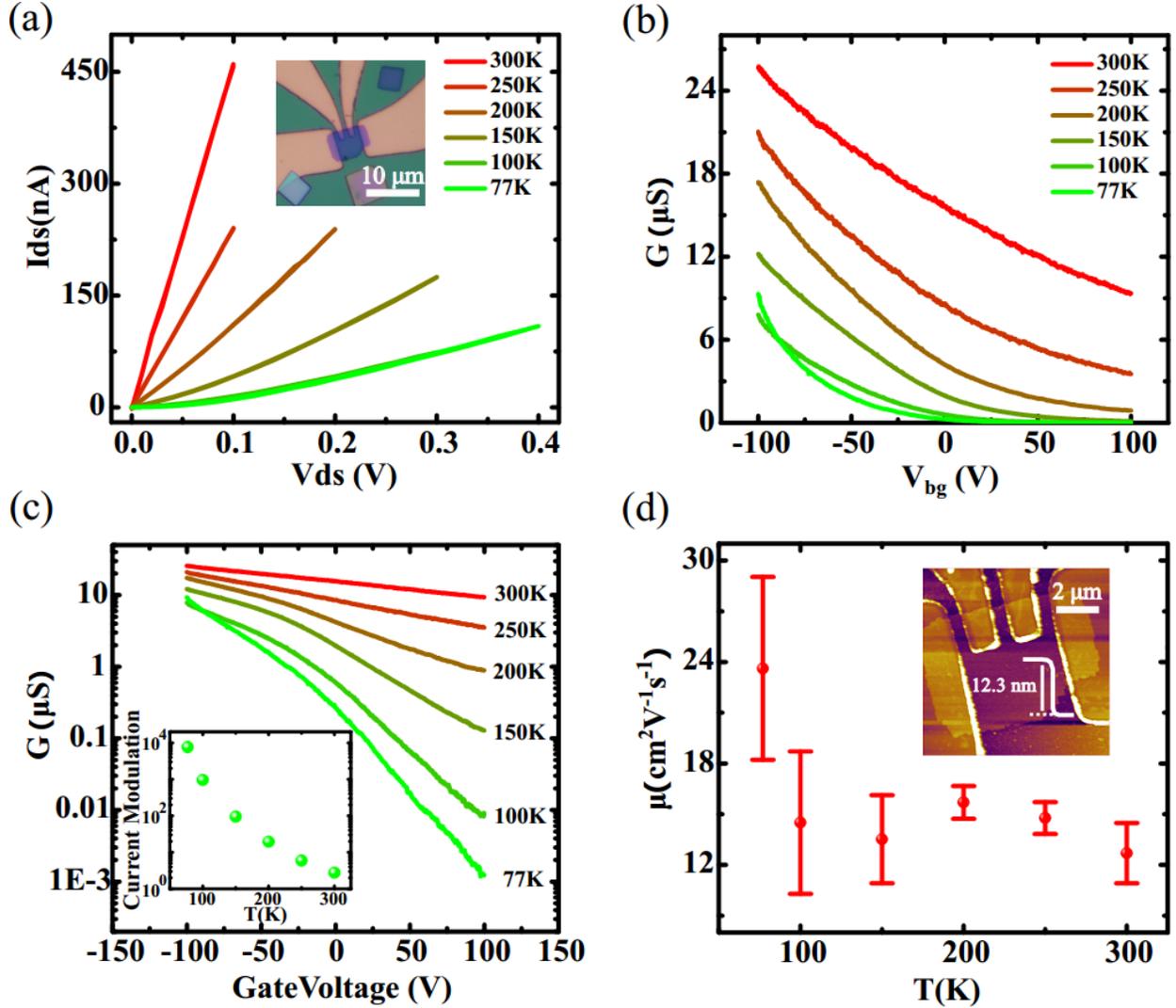

**Figure 3.** (a) $I_{ds}$ versus $V_{ds}$ at different temperatures ($T$) with the back gate grounded. Inset: an OM image of the FET device. (b) Four-probe conductance $G$ as a function of back gate voltage ($V_{bg}$) acquired at various $T$. (c) Semi-log scale plot of (b). Inset: the current modulation as a function of $T$. (d) Mobility $\mu$ at various $T$. Since the transfer curves in (b) are not linear, the error bars represent the maximum and minimum slopes used to calculate mobility with Equation 1. Inset: an AFM image of the device.

In order to study its electronic properties, we fabricated FET devices based on 2D SnS. Details of the device fabrication process can be found in the Methods. The OM and AFM images of a typical device with a 12.3 nm thick nanoplate on 300 nm $SiO_2$ are displayed in the insets of Figures 3a and 3d, respectively. Two-terminal *I-V* curves shown in Figure 3a evolve from linear to super linear as the device is cooled down, which indicates a contact barrier height comparable with the thermal energy at room temperature. To eliminate the effect of contact resistances, four-terminal conductance versus back-gate voltage ($V_{bg}$) measured at different temperatures is shown in Figure 3b, with the corresponding semi-log plot in Figure 3c. The synthesized 2D SnS shows a typical p type behavior. At room temperature (RT), the device cannot be turned off within the $V_{bg}$ range from -100 V to 100 V, indicating a heavy doping ($\sim 10^{19}$ $cm^{-3}$ at 300K). Theoretical calculation[16] of different intrinsic defects has shown that Sn vacancy has the lowest formation enthalpy ($\sim 0.8$ eV) and can act as a shallow acceptor with an energy level of about 42 meV from the valence band edge. As temperature goes down, the density of thermally excited holes decreases and gate voltage gains more control over the conduction of the channel. Below 100 K, the device can be turned off within the $V_{bg}$ range and the current modulation increases from about 3 at RT to $10^4$ at 77 K (see the inset of Figure 3c). From the data in Figure 3b, we calculate the hole mobility $\mu$ at different temperatures (shown in Figure 3d) with the equation[32]:

$$\mu = \frac{L}{W}\frac{dG}{C_g dV_g} = \frac{d\sigma}{C_g dV_g}, (1)$$

where *L* and *W* are the length and width of the channel, $C_g$ is the gate capacitance per unit area, *G* the conductance and $\sigma$ the conductivity. Mobility at RT is about 12 $cm^2V^{-1}s^{-1}$, and shows an overall increase at lower temperatures, in consistent with electron-phonon interaction being one of the main scattering mechanisms.[33] Theoretical calculation[28] has shown that the in-plane mobility of SnS with doping level of $10^{17}$ $cm^{-3}$ is about 100 $cm^2V^{-1}s^{-1}$, consistent with measured

RT mobility (~ 90 cm$^2$V$^{-1}$s$^{-1}$) of single crystal SnS.[34] This indicates a large room for improvement of the quality of our synthesized 2D SnS nanoplates.

With the puckered crystal lattice, 2D black phosphorus has very intriguing in-plane anisotropy[5-7] due to the different electronic structures along the armchair and zigzag directions. As an analogue to phosphorene, 2D SnS has also been predicted to have strong in-plane anisotropy.[28] To study this property, it is necessary to identify the two principle axes. Interestingly, the rhombic shape and well-defined straight edges of the as-grown SnS nanoplates make the identification a much easier task, compared with that of exfoliated 2D nanoplates. Through careful examination of the HRTEM and corresponding low magnification TEM images of the SnS nanoplates, we can identify that the edges of the nanoplates consist of {011} planes. One such example is shown in Figures 4a and 4b. Yellow and red arrows through the two figures demonstrate the alignments between atomic lattices and macroscopic crystal axes. The angle between the two neighboring edges in Figure 4b is also the same as the angle between [011] and [01$\bar{1}$] in Figure 4a. These results confirm that the shape of the nanoplates provides an easy way to distinguish between the armchair and zigzag directions, i.e. the longer diagonal is along the [010] (armchair) direction and the shorter one along the [001] (zigzag) direction. From crystal growth theory[35], the rhombic shape and the atomic direction assignments are not coincidences for a few nanoplates examined by the TEM. Rather they are results of the thermal dynamic process and apply to all the nanoplates that have reached equilibrium in their growth. In short, the {011} planes have the smallest density of dangling bonds and hence the lowest specific surface energy compared with other planes. So nanoplates with {011} terminated edges are thermodynamically the most stable (see Figure S1 for more details).

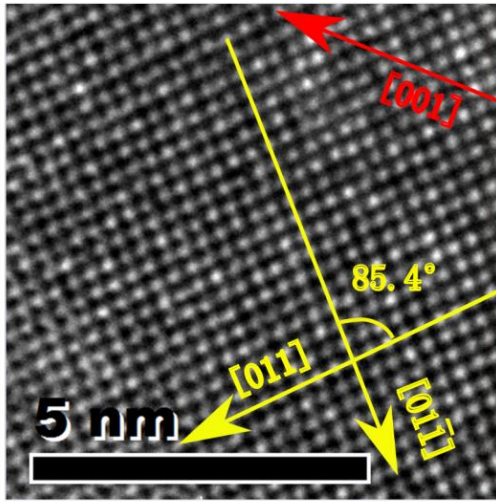
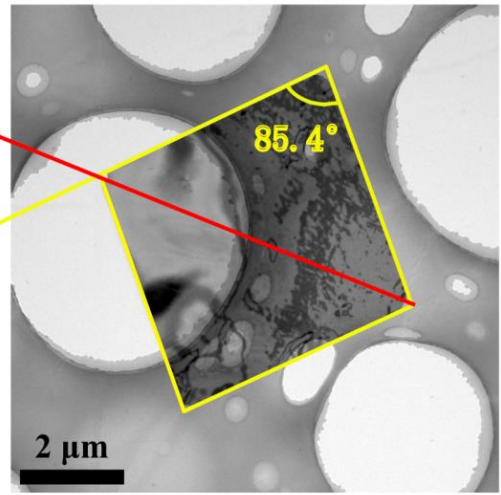
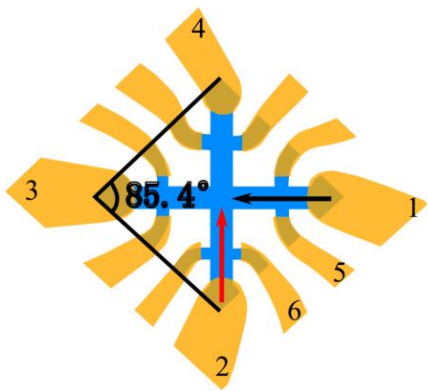
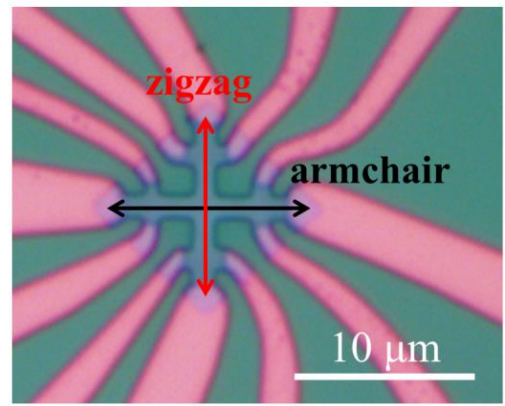
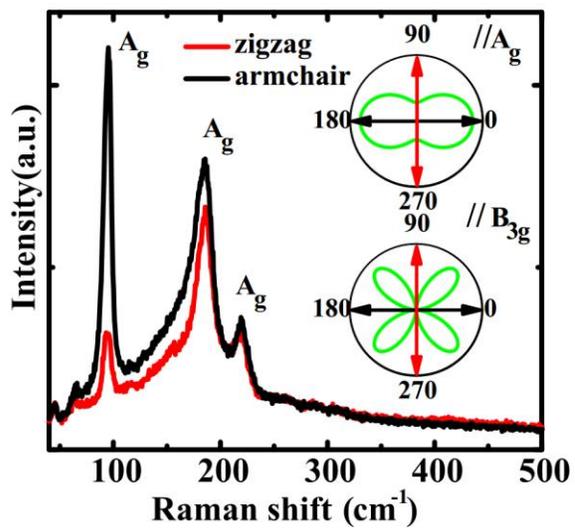
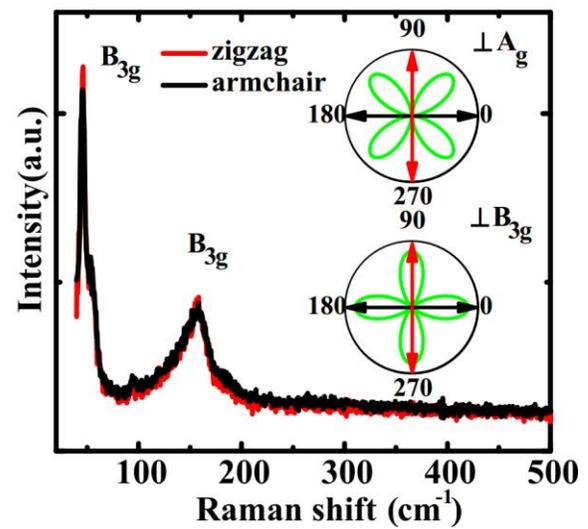

**Figure 4.** (a, b) HRTEM image and corresponding low magnification TEM image of a typical SnS nanoplate. The yellow and red arrows through the two figures are along the [011] and [001] (zigzag) directions. (c, d) Schematic and OM image of the 12.8 nm SnS FET device on 300 nm SiO$_2$. The black and red arrows in (c) indicate the anisotropic current-flow directions along armchair and zigzag directions, respectively. (e, f) Raman spectra of the SnS FET device shown in (d). The polarizations of the incident and Raman scattered light are parallel in (e) and perpendicular in (f). Insets: polar plots of Raman intensity as a function of the angle between the incident polarization and the armchair direction.

With this understanding, we designed multi-terminal FET devices as shown in Figure 4c to measure the electrical transport anisotropy of the as-grown SnS nanoplates. It is essentially a combination of two perpendicular Hall bars with current paths along the diagonals of the rhombus, which are also the zigzag or armchair directions as shown by the TEM study. Then EBL and H$_2$ plasma etching (Plasma-Preen Ⅱ-862) are used to fabricate the devices (Figure 4d). Since the nanoplates used to make the etched FETs cannot be examined by the TEM, polarized Raman spectroscopy is carried out to further confirm the assignments of the two crystal axes in the devices. Previous study[36] has shown that the intensity of A$_g$ mode Raman peaks can be used to determine the armchair and zigzag directions. In this technique, the intensity $I$ of a Raman peak can be expressed as $I \propto |e_i \cdot R \cdot e_s|^2$, where $e_i$ ($e_s$) is the polarization vector of incident (scattered) light and $R$ the Raman tensor.[37] Since only the A$_g$ and B$_{3g}$ modes are active in our measurement (see Figure 2d), we only need to consider their Raman tensors which can be found in tables for different point groups.[38] SnS belongs to the point group of D$_{2h}$, and the Raman tensors for A$_g$ and B$_{3g}$ modes in this group are

$$R(A_g) = \begin{pmatrix} A & 0 & 0 \\ 0 & B & 0 \\ 0 & 0 & C \end{pmatrix} \text{ and } R(B_{3g}) = \begin{pmatrix} 0 & 0 & 0 \\ 0 & 0 & D \\ 0 & D & 0 \end{pmatrix}.$$

The polarization vector for incident light is $e_i = (0, \cos\theta, \sin\theta)$, where $\theta$ is the angle between $e_i$ and the armchair direction of the SnS device. Similarly, $e_s$ can be expressed as $e_s = (0, \cos\theta, \sin\theta)$ or $(0, -\sin\theta, \cos\theta)$ under parallel or perpendicular polarization configurations, respectively. The intensities for different modes can be obtained as follows

$$I(A_g, \parallel) \propto (B\cos^2\theta + C\sin^2\theta)^2, \quad I(B_{3g}, \parallel) \propto D^2\sin^2 2\theta,$$

$$I(A_g, \perp) \propto \frac{(C-B)^2}{4}\sin^2 2\theta, \quad I(B_{3g}, \perp) \propto D^2\cos^2 2\theta,$$

where $\parallel$ ($\perp$) stands for parallel (perpendicular) polarization. Polar plots of these equations are shown as the insets of Figures 4e and 4f. It is interesting to note that when $\theta = 0°$ or $90°$, only the $A_g$ ($B_{3g}$) modes are selected under parallel (perpendicular) polarization. While the intensity of $B_{3g}$ modes should be the same for $\theta = 0°$ and $90°$, the intensity of the $A_g$ mode is maximized when $\theta = 0°$. So this technique can be used to identify the armchair and zigzag directions. Measurements are performed on the device of Figure 4d. Raman spectra shown in Figures 4e and 4f agree very well with the theoretical prediction. This result corroborates our TEM analysis and validates the assignment of the armchair and zigzag directions in functional FET devices. It lays a solid foundation for the following study of electrical transport anisotropy.

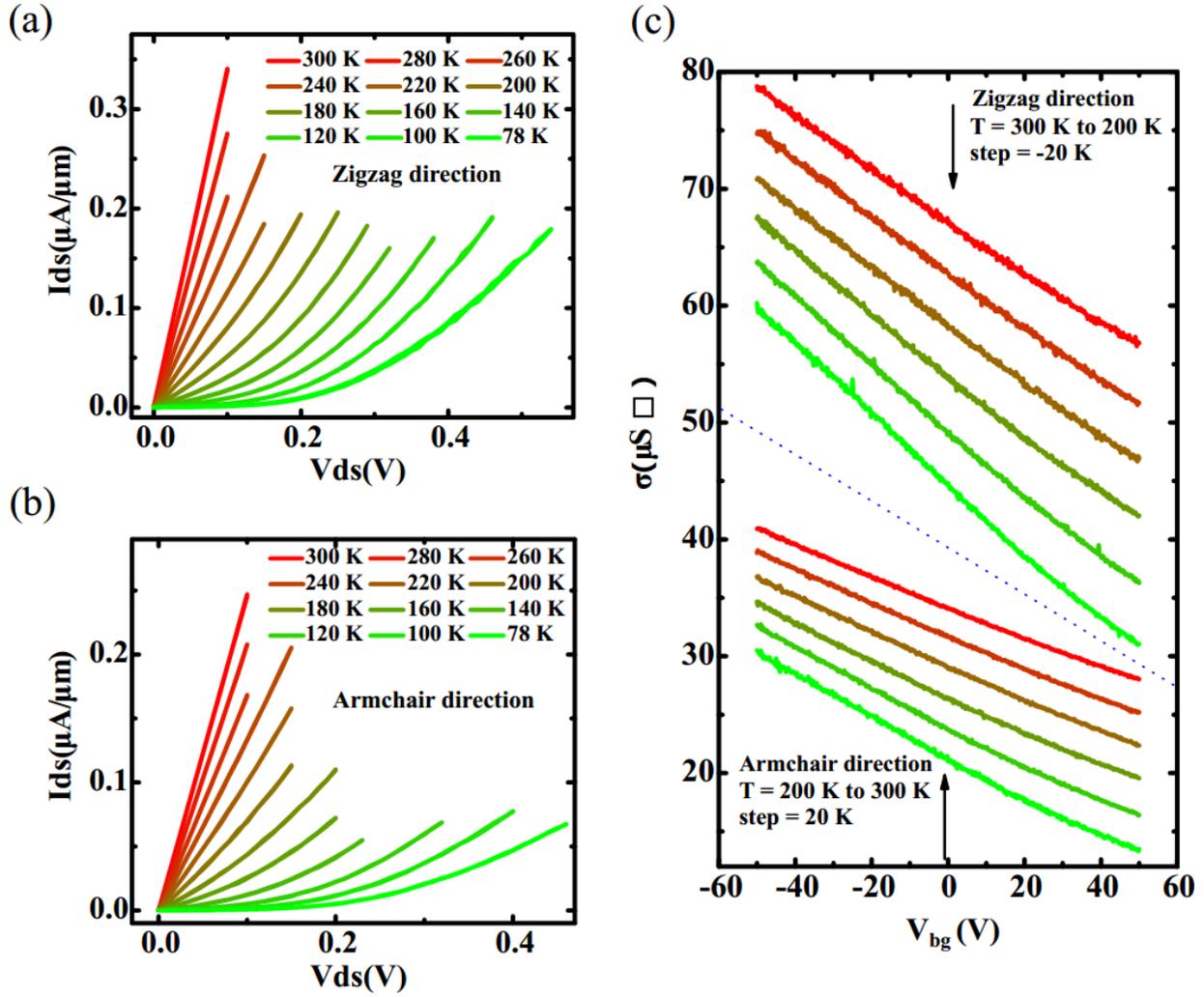

**Figure 5.** (a, b) The $I_{ds}$-$V_{ds}$ plot of the device shown in Figure 4d at various $T$ along the zigzag (a) and armchair (b) directions with the back gate grounded. (c) Transfer characteristic at different $T$ along the zigzag (upper part) and armchair (lower part) directions. Data are plotted in their original values; no vertical shift is applied. The dotted line is added to guide the eye.

The electrical anisotropy is studied based on devices with the cross-Hall-bar structure shown in Figure 4c. In total three such devices were measured and they all presented similar anisotropic behavior. The results of one such device are presented in Figure 5 (see Figure S2 for the characterization of another device). Temperature-dependent two-terminal $I$-$V$ curves along the zigzag and armchair directions are plotted in Figures 5a and b, respectively. In previous studies on the anisotropy of black phosphorus[3, 7] and ReS$_2$,[39] the 2D nanoplates were etched into circular shape, and angle-dependent two-terminal conductance was measured. The applicability of this method is based on the assumption that all the metal-2D semiconductor contacts have the same resistance. However, the Schottky barrier heights are often different from one contact to another, even within the same device.[40] From Figures 5a and 5b, it can be seen that current injection along the zigzag and armchair directions has different temperature dependence, indicating different barrier heights which can be detrimental to an accurate measurement of the intrinsic anisotropy. On the other hand, the cross-Hall-bar device structure of Figure 4c allows us to perform four-terminal characterization along the two directions on a same device without the contact complication. In Figure 5c, the temperature-dependent transfer curves along the zigzag and armchair directions are plotted together. The curves are naturally segregated into two groups, and a dotted line is added to guide the eye. The group on the top always has higher conductivities and larger slopes, indicating higher carrier mobility along the zigzag direction. It is interesting to note that in phosphorene the zigzag direction is the one who has the lower mobility.[7]

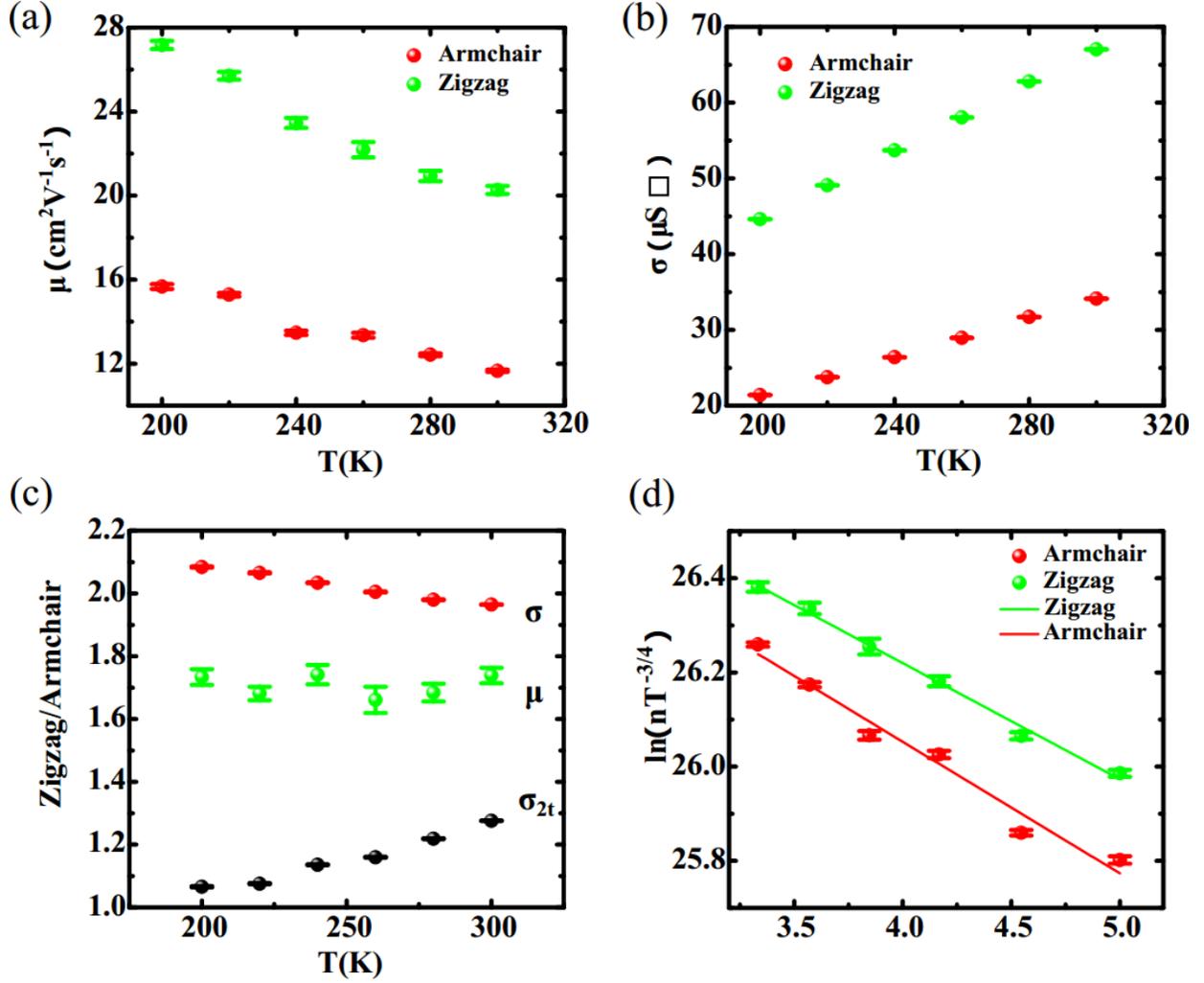

**Figure 6.** (a) The mobility $\mu$ extracted from Figure 5c at zero gate voltage. (b) The four-terminal sheet conductivity versus $T$ obtained from Figure 5c, also at zero gate voltage. (c) The temperature-dependent ratio of mobility $\mu$, four-terminal sheet conductivity $\sigma$ and two-terminal conductivity $\sigma_{2t}$ between the zigzag and armchair directions. It can be seen that two-terminal measurements severely underestimate the intrinsic anisotropy. (d) Carrier density as a function of temperature along the two directions is used to extract the defect energy level. $\ln nT^{-3/4}$ is plotted against $1000/T$, where the units of $n$ and $T$ are cm$^{-2}$ and K, respectively. The solid lines are linear fits of the data, which give a defect energy level of ~ 45 meV.

To quantify these observations, temperature-dependent mobility along the zigzag and armchair directions are calculated with Equation 1 and plotted in Figure 6a. It is obvious to see that the zigzag direction always has a higher mobility. Since $\mu = e\tau/m$, the anisotropy can be explained by the different effective masses along the two principle axes. Theoretical calculation[28] has shown that the effective mass ratio of these two directions is $m_{zigzag}/m_{armchair} = 0.21/0.36$, which gives a mobility ratio of $\mu_{zigzag}/\mu_{armchair} \approx 1.7$. To compare with the prediction, we plot the mobility ratio in Figure 6c, which shows excellent agreement with the theory and has little temperature dependence. Since the conductivity can be expressed as $\sigma = n\mu e$, we expect $\sigma$ to have the same anisotropy as $\mu$. Indeed, as shown in Figure 6b $\sigma$ along the zigzag direction is always higher than that along the armchair direction; and the ratio is plotted in Figure 6c as well. In principle, the ratios of $\sigma$ and $\mu$ should be identical, if carrier densities along the two arms of the cross Hall bar are the same. Here the difference between the two ratios may come from spatially varying doping densities ($n \approx 1.83 \times 10^{13}$ cm$^{-2}$ and $2.06 \times 10^{13}$ cm$^{-2}$ along armchair and zigzag direction respectively at 300K) within the device. Similar variation has been observed in 2D materials such as graphene,[41] where electron-hole puddles could be formed. In the Supporting Information (Figure S3), another device measured in this study shows similar anisotropic ratios in $\sigma$ and $\mu$, presumably due to a more homogeneous doping. In Figure 6c, also plotted is the two-terminal conductivity ratio along the two axes with data from Figures 5a and 5b. Obviously it severely underestimates the intrinsic anisotropy due the contact issue discussed above.

With the data of $\sigma$ and $\mu$ in Figures 6a and b, we can obtain the carrier density as a function of temperature, $n(T)$, from which the acceptor energy level can be extracted (Figure 6d). From semiconductor physics, we have:[42, 43]

$$n(T) = \left(\frac{N_A N_V}{2}\right)^{1/2} e^{-E_a/2kT}, \qquad N_V = 2\left(\frac{2\pi m_{dh} kT}{h^2}\right)^{3/2}, \qquad E_a = E_V - E_A$$

where $N_A$ is the acceptor density, $N_V$ the density of states near the valence band maximum (VBM), $E_a$ the activation energy, $E_V$ the energy level of the VBM, $E_A$ the acceptor energy level, $k$ the Bolzmann constant, $m_{dh}$ the density-of-state effective mass of the VBM and $h$ the Planck constant. Since $N_A$ is a constant for a specific sample, $\ln n T^{-3/4}$ is linearly proportional to $-E_a/2kT$. Activation energy $E_a$ can be extracted from the linear fits in Figure 6d, which equals to $46.5 \pm 1.7$ meV along the armchair direction and $43.1 \pm 0.8$ meV along the zigzag direction. These measured values confirm that the major dopants are shallow Sn vacancies, whose acceptor energy level was predicted[16] to be 42 meV above the VBM.

CONCLUSIONS

In summary, we have successfully grown rhombic 2D SnS nanoplates on mica with lateral dimension 5-15 μm and thickness 6-20 nm using the PVT method. Hole mobility exceeding 20 $cm^2V^{-1}s^{-1}$ and current on-off ratio up to $10^4$ at 77K were obtained from the as-grown SnS. The long and short diagonals of the rhombus SnS nanoplates were found to be along the [010] (armchair) and [001] (zigzag) directions with the help of the TEM and polarized Raman spectroscopy. FETs with cross-Hall-bar structure were made to study the in-plane anisotropic transport properties. Mobility along the zigzag direction was measured to be up to 1.7 times of that in the armchair direction. Finally, activation energy of ~ 45 meV was extracted from the temperature-dependent carrier density, which confirms Sn vacancies as the major dopants when compared with theoretical calculation. Our study is one of the first experimental efforts in exploring the intriguing properties of 2D SnS and paved the way for future study of 2D monochalcogenides.

METHODS

**Synthesis of 2D SnS nanoplates.** The 2D SnS was synthesized in a horizontal double-zone tube furnace (AnHui BEQ Equipment Technology CO., Ltd.) with a 2 inch diameter quartz tube. The schematic is shown in Figure 1a. SnS powder (99.5％, Alfa Aesar) in a quartz boat placed in the center of the high temperature zone was adopted as the source material. Freshly cleaved mica ($KMg_3(AlSi_3O_{10})F_2$, Changchun Taiyuan Fluorphlogopite CO., Ltd.) served as the substrate were placed on a glass slide after the SnS source (the distance between source and slide is about 12 cm). Ultra-pure argon flew in the quartz tube from the high temperature side through a gas mass flow controller (Alicat Scientific, MC-200SCCM-D/5M) with a flow rate of 200 SCCM, and was evacuated from the low temperature side with a mechanical pump. Before starting the growth, the quartz tube was flushed with ultra-pure argon three times to avoid oxygen contamination. Then the temperatures of the two heating zones were raised to 530 ℃~560 ℃ and 350 ℃ respectively in 50 minutes, and maintained these temperatures for one hour with the growth pressure at about 250 Pa. Finally, the furnace was cooled down to room temperature naturally.

**Raman Spectroscopy.** The Raman spectra in Figure 2 were obtained with a Raman microscopy system (Thermo Fisher Scientific, DXR) with 532 nm elliptically polarized laser. No particular polarization was selected from the Raman scattered light. The polarized Raman spectra shown in Figure 4 were obtained with Jobin Yvon HR-Evolution 2, where a 532 nm linearly polarized laser was used as the excitation and an adjustable polarizer was added at the entrance of the spectrometer.

**Device Fabrication and Characterization.** The electrodes of the SnS FET devices were fabricated by e-beam lithography (Carl Zeiss S-4700 SEM and Raith). 7 nm Ti and 30 nm Au were deposited by thermal evaporation (Angstrom Engineering). Temperature-dependent electrical measurements were done at vacuum ($2 \times 10^{-3}$ Pa) and in the dark with a Janis ST-500-1 probe station using Keithley 2612B SourceMeters.

**Supporting Information Available.** Figure S1 illustrates the density of dangling bonds and hence the specific surface energy along different edges of 2D SnS nanoplates. Figure S2 and S3 are the electrical characterization of a 13.3 nm SnS device with cross-Hall-bar structure.

AUTHOR INFORMATION

**Corresponding Author**

*xuejm@shanghaitech.edu.cn.

**Author Contributions**

The manuscript was written through contributions of all authors. All authors have given approval to the final version of the manuscript.

**Funding Sources**

This work was supported by National Natural Science Foundation of China (Grant No. 11504234), Science and Technology Commission of Shanghai Municipality (Grant No. 15QA1403200) and ShanghaiTech University. The nanofabrication facility was supported by the Strategic Priority Research Program (B) of the Chinese Academy of Sciences (Grant No. XDB04030000).

ACKNOWLEDGMENT

We thank Professor Hung-Ta Wang of ShanghaiTech University for his great help with the manuscript. This work was supported by National Natural Science Foundation of China (Grant No. 11504234), Science and Technology Commission of Shanghai Municipality (Grant No. 15QA1403200) and ShanghaiTech University. The nanofabrication facility was supported by the Strategic Priority Research Program (B) of the Chinese Academy of Sciences (Grant No. XDB04030000).

REFERENCES

1. Li, L.; Yu, Y.; Ye, G. J.; Ge, Q.; Ou, X.; Wu, H.; Feng, D.; Chen, X. H.; Zhang, Y. Black phosphorus field-effect transistors. *Nature nanotechnology* **2014,** *9* (5), 372-377.

2. Kim, J.; Baik, S. S.; Ryu, S. H.; Sohn, Y.; Park, S.; Park, B. G.; Denlinger, J.; Yi, Y.; Choi, H. J.; Kim, K. S. 2D MATERIALS. Observation of tunable band gap and anisotropic Dirac semimetal state in black phosphorus. *Science* **2015,** *349* (6249), 723-6.

3. Liu, H.; Neal, A. T.; Zhu, Z.; Luo, Z.; Xu, X.; Tomanek, D.; Ye, P. D. Phosphorene: an unexplored 2D semiconductor with a high hole mobility. *ACS nano* **2014,** *8* (4), 4033-41.

4. Zhang, S.; Yang, J.; Xu, R.; Wang, F.; Li, W.; Ghufran, M.; Zhang, Y. W.; Yu, Z.; Zhang, G.; Qin, Q.; Lu, Y. Extraordinary photoluminescence and strong temperature/angle-dependent Raman responses in few-layer phosphorene. *ACS nano* **2014,** *8* (9), 9590-6.

5. Phaneuf-L'Heureux, A.-L.; Favron, A.; Germain, J.-F.; Lavoie, P.; Desjardins, P.; Leonelli, R.; Martel, R.; Francoeur, S. Polarization-Resolved Raman Study of Bulk-like and Davydov-Induced Vibrational Modes of Exfoliated Black Phosphorus. *Nano Letters* **2016**.

6. He, J.; He, D.; Wang, Y.; Cui, Q.; Bellus, M. Z.; Chiu, H. Y.; Zhao, H. Exceptional and Anisotropic Transport Properties of Photocarriers in Black Phosphorus. *ACS nano* **2015,** *9* (6), 6436-42.

7. Xia, F.; Wang, H.; Jia, Y. Rediscovering black phosphorus as an anisotropic layered material for optoelectronics and electronics. *Nature communications* **2014,** *5*, 4458.

8. Steinmann, V.; Jaramillo, R.; Hartman, K.; Chakraborty, R.; Brandt, R. E.; Poindexter, J. R.; Lee, Y. S.; Sun, L.; Polizzotti, A.; Park, H. H.; Gordon, R. G.; Buonassisi, T. 3.88% efficient tin sulfide solar cells using congruent thermal evaporation. *Adv Mater* **2014,** *26* (44), 7488-92.

9. Sinsermsuksakul, P.; Hartman, K.; Bok Kim, S.; Heo, J.; Sun, L.; Hejin Park, H.; Chakraborty, R.; Buonassisi, T.; Gordon, R. G. Enhancing the efficiency of SnS solar cells via band-offset engineering with a zinc oxysulfide buffer layer. *Applied Physics Letters* **2013,** *102* (5), 053901.

10. Sinsermsuksakul, P.; Sun, L.; Lee, S. W.; Park, H. H.; Kim, S. B.; Yang, C.; Gordon, R. G. Overcoming Efficiency Limitations of SnS-Based Solar Cells. *Advanced Energy Materials* **2014,** *4* (15), 1400496.

11. Rodin, A. S.; Gomes, L. C.; Carvalho, A.; Castro Neto, A. H. Valley physics in tin (II) sulfide. *Physical Review B* **2016,** *93* (4).

12. Fei, R.; Li, W.; Li, J.; Yang, L. Giant piezoelectricity of monolayer group IV monochalcogenides: SnSe, SnS, GeSe, and GeS. *Applied Physics Letters* **2015,** *107* (17), 173104.

13. Hanakata, P. Z.; Carvalho, A.; Campbell, D. K.; Park, H. S. Polarization and valley switching in monolayer group-IV monochalcogenides. *Physical Review B* **2016,** *94* (3).

14. Wu, M.; Zeng, X. C. Intrinsic Ferroelasticity and/or Multiferroicity in Two-Dimensional Phosphorene and Phosphorene Analogues. *Nano Lett* **2016,** *16* (5), 3236-41.

15. Li, S. L.; Tsukagoshi, K.; Orgiu, E.; Samori, P. Charge transport and mobility engineering in two-dimensional transition metal chalcogenide semiconductors. *Chem Soc Rev* **2016,** *45* (1), 118-51.

16. Vidal, J.; Lany, S.; d'Avezac, M.; Zunger, A.; Zakutayev, A.; Francis, J.; Tate, J. Band-structure, optical properties, and defect physics of the photovoltaic semiconductor SnS. *Applied Physics Letters* **2012,** *100* (3), 032104.

17. Zhao, S.; Wang, H.; Zhou, Y.; Liao, L.; Jiang, Y.; Yang, X.; Chen, G.; Lin, M.; Wang, Y.; Peng, H.; Liu, Z. Controlled synthesis of single-crystal SnSe nanoplates. *Nano Research* **2015,** *8* (1), 288-295.

18. Noguchi, H.; Setiyadi, A.; Tanamura, H.; Nagatomo, T.; Omoto, O. Characterization of vacuum-evaporated tin sulfide film for solar cell materials. *Solar energy materials and solar cells* **1994,** *35*, 325-331.

19. Johnson, J.; Jones, H.; Latham, B.; Parker, J.; Engelken, R.; Barber, C. Optimization of photoconductivity in vacuum-evaporated tin sulfide thin films. *Semiconductor Science and Technology* **1999,** *14* (6), 501.

20. Reina, A.; Son, H.; Jiao, L.; Fan, B.; Dresselhaus, M. S.; Liu, Z.; Kong, J. Transferring and identification of single-and few-layer graphene on arbitrary substrates. *The Journal of Physical Chemistry C* **2008,** *112* (46), 17741-17744.

21. Liang, X.; Sperling, B. A.; Calizo, I.; Cheng, G.; Hacker, C. A.; Zhang, Q.; Obeng, Y.; Yan, K.; Peng, H.; Li, Q. Toward clean and crackless transfer of graphene. *ACS nano* **2011,** *5* (11), 9144-9153.

22. Li, X.; Zhu, Y.; Cai, W.; Borysiak, M.; Han, B.; Chen, D.; Piner, R. D.; Colombo, L.; Ruoff, R. S. Transfer of large-area graphene films for high-performance transparent conductive electrodes. *Nano letters* **2009,** *9* (12), 4359-4363.

23. Guo, L.; Yan, H.; Moore, Q.; Buettner, M.; Song, J.; Li, L.; Araujo, P. T.; Wang, H. T. Elastic properties of van der Waals epitaxy grown bismuth telluride 2D nanosheets. *Nanoscale* **2015,** *7* (28), 11915-21.

24. Li, H.; Wu, J. M. T.; Huang, X.; Yin, Z. Y.; Liu, J. Q.; Zhang, H. A Universal, Rapid Method for Clean Transfer of Nanostructures onto Various Substrates. *ACS nano* **2014,** *8* (7), 6563-6570.

25. Castellanos-Gomez, A. Black Phosphorus: Narrow Gap, Wide Applications. *J Phys Chem Lett* **2015,** *6* (21), 4280-91.

26. Materials Project. https://www.materialsproject.org/materials/mp-2231/.

27. Madelung, O. *Semiconductors: Data Handbook*. Third Edition ed.; HARBIN INSTITUTE OF TECHNOLOGY PRESS: 2014.

28. Guo, R.; Wang, X.; Kuang, Y.; Huang, B. First-principles study of anisotropic thermoelectric transport properties of IV-VI semiconductor compounds SnSe and SnS. *Physical Review B* **2015,** *92* (11).

29. Wangperawong, A.; Herron, S. M.; Runser, R. R.; Hägglund, C.; Tanskanen, J. T.; Lee, H.-B.-R.; Clemens, B. M.; Bent, S. F. Vapor transport deposition and epitaxy of orthorhombic SnS on glass and NaCl substrates. *Applied Physics Letters* **2013,** *103* (5), 052105.

30. Nikolic, P.; Mihajlovic, P.; Lavrencic, B. Splitting and coupling of lattice modes in the layer compound SnS. *Journal of Physics C: Solid State Physics* **1977,** *10* (11), L289.

31. Chandrasekhar, H. R.; Humphreys, R. G.; Zwick, U.; Cardona, M. Infrared and Raman spectra of the IV-VI compounds SnS and SnSe. *Physical Review B* **1977,** *15* (4), 2177-2183.

32. Schroder, D. K. *Semiconductor material and device characterization*. Third Edition ed.; A Wiley-Interscience Publication: 2006; p 500-502.

33. Kim, S.; Konar, A.; Hwang, W. S.; Lee, J. H.; Lee, J.; Yang, J.; Jung, C.; Kim, H.; Yoo, J. B.; Choi, J. Y.; Jin, Y. W.; Lee, S. Y.; Jena, D.; Choi, W.; Kim, K. High-mobility and low-power thin-film transistors based on multilayer MoS2 crystals. *Nature communications* **2012,** *3*, 1011.

34. Albers, W.; Haas, C.; Vink, H. J.; Wasscher, J. D. Investigations on SnS. *Journal of Applied Physics* **1961,** *32* (10), 2220.

35. Markov, I. V. *Crystal growth for beginners: fundermentals of nucleation, crystal growth, and epitaxy.* 2nd Edition ed.; World Scientific Publishing Co.Pte.Ltd.: 2004; p 14.

36. Xia, J.; Li, X. Z.; Huang, X.; Mao, N.; Zhu, D. D.; Wang, L.; Xu, H.; Meng, X. M. Physical vapor deposition synthesis of two-dimensional orthorhombic SnS nanoplates with strong angle/temperature-dependent Raman responses. *Nanoscale* **2016,** *8* (4), 2063-70.

37. Garmire, E.; Hammer, J.; Kogelnik, H.; Tamir, T.; Zernike, F.; Cardona, M. *Light Scattering in Solids 1*. Springer Science & Business Media: 2013; Vol. 8.

38. Raman and Hyper-Raman Tensors. http://www.cryst.ehu.es/cryst/transformtensor.html.

39. Liu, E.; Fu, Y.; Wang, Y.; Feng, Y.; Liu, H.; Wan, X.; Zhou, W.; Wang, B.; Shao, L.; Ho, C. H.; Huang, Y. S.; Cao, Z.; Wang, L.; Li, A.; Zeng, J.; Song, F.; Wang, X.; Shi, Y.; Yuan, H.; Hwang, H. Y., et al. Integrated digital inverters based on two-dimensional anisotropic ReS2 field-effect transistors. *Nature communications* **2015,** *6*, 6991.

40. Larentis, S.; Fallahazad, B.; Tutuc, E. Field-effect transistors and intrinsic mobility in ultra-thin MoSe2 layers. *Applied Physics Letters* **2012,** *101* (22), 223104.

41. Martin, J.; Akerman, N.; Ulbricht, G.; Lohmann, T.; Smet, J. H.; von Klitzing, K.; Yacoby, A. Observation of electron–hole puddles in graphene using a scanning single-electron transistor. *Nature Physics* **2007,** *4* (2), 144-148.

42. B.I.Shklovskii; A.L.Efros. *Electronic properties of doped semiconductor*. Springer: 1984; p 75.

43. S.M.Sze; K.Ng, K. *Physics of semiconductor devices*. A Wiley-Interscience Publication: 2007; p 24-25.